\documentclass[12pt]{article}

\usepackage{amssymb}
\usepackage{amstext}

\def\be{\begin{equation}}

\def\ee{\end{equation}}

\def\ba{\begin{eqnarray}}

\def\ea{\end{eqnarray}}

\def \mul {\odot}

\def \ba {\bar}

\def \2 {{1 \over 2}}
\def \3 {{1 \over 3}}
\def \4 {{1 \over 4}}
\def \5 {{1 \over 5}}
\def \6 {{1 \over 6}}
\def \7 {{1 \over 7}}
\def \8 {{1 \over 8}}
\def \9 {{1 \over 9}}
\def \0 { \infty}

\def \qq {\qquad}

\begin{document}
\pagestyle{empty}
\begin{flushright}
\begin{tabular}{ll}
QMUL-PH-02-15 & \\
hep-th/0208155 & \\
August 2002 & \\ [.5in]
\end{tabular}
\end{flushright}
\begin{center}
{\Large {\bf{Exotic tensor gauge theory and duality}}} \\ [.7in]
{\large{P. de Medeiros and C. Hull}} \\ [.3in]
{\emph{Physics Department, Queen Mary, University of London, \\ Mile End Road, London E1 4NS, U.K.}} \\
{\tt{p.demedeiros@qmul.ac.uk}} , {\tt{c.m.hull@qmul.ac.uk}} \\ [.9in]
{\large{\bf{Abstract}}} \\ [.2in]
\end{center}
Gauge fields in exotic representations of the Lorentz group in $D$ dimensions -- i.e. ones which are tensors of mixed symmetry corresponding to Young  tableaux with arbitrary numbers of rows and columns -- naturally arise through massive string modes and in dualising gravity and other theories in higher dimensions. We generalise the formalism of differential forms to allow the discussion of arbitrary gauge fields. We present the gauge symmetries, field strengths, field equations and actions for the free theory, and construct the various dual theories. In particular, we discuss linearised gravity in arbitrary dimensions, and its two dual forms.
\clearpage
\pagestyle{plain}
\pagenumbering{arabic}
\line(1,0){475}      
\tableofcontents
\line(1,0){475}


\section{Introduction}
Fields in  symmetric or antisymmetric tensor representations of the Lorentz group occur in many contexts. However, tensor fields in more exotic representations corresponding to arbitrary Young tableaux can also occur. Such fields arise among the higher-spin massive modes in string theory, and can also occur in dualising some of the more familiar tensor fields. They arise too in higher spin gauge theories in dimensions $5 \leq D \leq 7$ {\cite{keysezsun}}. The free covariant field theories for such representations have been discussed in {\cite{keylabmor}}, {\cite{keysieg}}, {\cite{keytsul}}. It is well known that a massless $p$-form gauge field $A$ in $D$ dimensions can be dualised to a $D-p-2$ form gauge field $B$, with $dA=*dB$, exchanging field equations with Bianchi identitites. This can be extended to other tensor fields. For example, a massive spin-two field in four dimensions, usually described in terms of a symmetric second rank tensor, has a dual description in terms of a third rank tensor $d_{ \mu\nu \rho}=d_{[ \mu\nu ]\rho}$ satisfying $d_{\mu\nu \rho} = - d_{ \nu\mu \rho}$ and $d_{[ \mu\nu\rho ]} =0$ {\cite{key1}}. It is then of mixed symmetry, corresponding to a Young tableau with two columns of length two and one. In {\cite{key2}}, {\cite{key3}}, {\cite{key4}}, {\cite{key5}}, dual forms of linearised gravity were found in arbitrary dimensions.

For example, in $D=5$, linearised gravity is formulated in the usual way in terms of a symmetric tensor $h_{\mu\nu}$. There are however two dual forms of this theory, one formulated in terms of a gauge field $d_{\mu\nu\rho} $ (again satisfying  $d_{\mu\nu \rho} = - d_{ \nu\mu \rho}$ and $d_{[ \mu\nu\rho ]} =0$) and one formulated in terms of a gauge field $c_{\mu\nu\rho\sigma}$ with the same symmetries as the Riemann tensor. These fields transform under the gauge transformations
%
%
\begin{equation}
\begin{array}{rcl}
\delta c_{\mu\nu\rho\sigma} &=& 2\, \partial_{[\mu} \chi_{\nu ] \rho\sigma} + 2\, \partial_{[\rho}
\chi_{\sigma ] \mu\nu} - 4\, \partial_{[\mu} \chi_{\nu\rho\sigma ]} \\
\delta d_{\mu\nu\rho} &=& 2\, \partial_{[\mu} \alpha_{\nu ] \rho} - 2\, \partial_{[\mu}
\alpha_{\nu\rho ]} + \partial_{\rho} \beta_{\mu\nu} - \partial_{[ \rho} \beta_{\mu\nu ]} \\
\delta h_{\mu\nu} &=& 2\, \partial_{( \mu} \xi_{\nu )}
\end{array} 
\label{eq:1}
\end{equation}
respectively, with parameters $\chi_{\mu [\nu \rho ]} = \chi_{\mu\nu\rho}$, $\alpha_{\mu\nu}$, $\beta_{[ \mu\nu ]} = \beta_{\mu\nu}$ and $\xi_{\mu}$, so that each gauge field corresponds to five physical degrees of freedom; in each case, these are in the {\bf{5}} representation of the little group $SO(3)$. The respective gauge invariant field strengths are
%
%
\begin{equation} G_{\mu\nu\rho\sigma\alpha\beta} \;\; = \;\; 9\, \partial_{[\mu} c_{\nu\rho ][\sigma\alpha ,\beta ]} , \quad S_{\mu\nu\rho\alpha\beta} \;\; = \;\; -6 \, \partial_{[\mu} d_{\nu\rho ][\alpha , \beta ]} , \quad R_{\mu\nu\rho\sigma} \;\; = \;\; -4 \, \partial_{[\mu} h_{\nu ][\rho ,\sigma ]}
\label{eq:2}
\end{equation}
which all involve two derivatives. In gravity, the linearised curvature satisfies the Bianchi identity
%
%
\begin{equation}
R_{[\mu\nu\rho]\sigma} \;\; =\;\; 0
\label{eq:3}
\end{equation}
and the linearised field equation
%
%
\begin{equation} 
\eta^{\mu\rho} R_{\mu\nu\rho\sigma} \;\; =\;\; 0
\label{eq:4}
\end{equation}
where $\eta_{\mu\nu}$ is the $SO(4,1)$-invariant Minkowski metric. The gauge field $d_{\mu\nu\rho}$ can be defined non-locally in terms of the gauge field $h_{\mu\nu}$ via the duality relation
%
%
\begin{equation}
S_{\mu\nu\rho\alpha\beta} \;\; = \;\; \frac{1}{2} \epsilon_{\mu\nu\rho\gamma\delta} \, R^{\gamma\delta}_{\;\;\;\; \alpha\beta} 
\label{eq:5}
\end{equation}
The gravitational field equation then implies that $S_{\mu\nu\rho\alpha\beta}$ satisfies the Bianchi identity
%
%
\begin{equation}
S_{[\mu\nu\rho\alpha]\beta} \;\; = \;\; 0
\label{eq:6}
\end{equation}
while the gravitational Bianchi identity implies the field equation
%
%
\begin{equation}
\eta^{\mu\alpha} S_{\mu\nu\rho\alpha\beta} \;\; = \;\; 0 
\label{eq:7}
\end{equation}
Similarly, one can introduce the gauge field $c_{\mu\nu\rho\sigma}$ via a further duality 
%
%
\begin{equation}
G_{\mu\nu\rho\sigma\alpha\beta} \;\; = \;\; \frac{1}{2} \epsilon_{\sigma\alpha\beta\gamma\delta} S_{\mu\nu\rho}^{\;\;\;\;\;\; \gamma\delta} 
\label{eq:8}
\end{equation}
whose field strength then satisfies the Bianchi identity
%
%
\begin{equation}
G_{[\mu\nu\rho\sigma]\alpha\beta} \;\; = \;\;  0
\label{eq:9}
\end{equation}
corresponding to the gravitational Bianchi identity and the field eqautions for $d_{\mu\nu\rho}$, and the field equation
%
%
\begin{equation}
\eta^{\mu\sigma} \eta^{\nu\alpha} G_{\mu\nu\rho\sigma\alpha\beta} \;\; = \;\;  0
\label{eq:10}
\end{equation}
corresponding to the gravitational field equation and the Bianchi identity for $S_{\mu\nu\rho\alpha\beta}$.

These dual forms of gravity in $D=5$ arise naturally from considering the reduction of the (4,0) supersymmetric free theory in $D=6$, in which the five physical degrees of freedom of the graviton in $D=5$ arise from the reduction of a gauge field $C^{+}_{MNPQ}$ in six dimensions with the symmetries of the Riemnann tensor and with field strength satisfying $D=6$ self-duality constraints {\cite{key2}}, {\cite{key4}}, {\cite{key5}}. In {\cite{key2}}, it was argued that M-theory compactified to $D=5$, with $D=5$, $N=8$ supergravity as the low-energy effective field theory, could have a strong-coupling limit giving rise to a $D=6$, (4,0) supersymmetric theory with gravity described by the exotic gauge field $C^{+}_{MNPQ}$.

This generalises to $D$ dimensions, where the graviton $h_{\mu\nu}$ is dual to a gauge field $D_{\mu_1...\mu_{D-3} \nu}$ (corresponding to a Young tableau with one column of length $D-3$ and one of length one) or to a gauge field $C_{\mu_1 ... \mu_{D-3} \nu_1 ... \nu_{D-3}}$ (corresponding to a Young tableau with two columns, both of length $D-3$).

Antisymmetric tensor gauge fields are naturally formulated in the language of differential forms. The purpose of this paper is to develop the corresponding formalism for these higher rank gauge fields and their dualities, following the work of {\cite{key7}}, {\cite{key8}}, {\cite{key9}}. Gauge fields such as $h_{\mu\nu}$, $D_{\mu_1...\mu_{D-3} \nu}$ or $C_{\mu_1 ... \mu_{D-3} \nu_1 ... \nu_{D-3}}$ represented by Young tableaux with two columns of length $p,q$ are elegantly described in terms of {\textit{bi-forms}}, taking values in the tensor product space $\Lambda^p \otimes \Lambda^q$ of $p$-forms with $q$-forms. This formalism is developed in section 2 and applied to gauge theories. The construction is generalised in section 3, using {\textit{multi-forms}} to establish dual descriptions of theories with gauge fields in arbitrary
representations of $GL(D,{\mathbb{R}})$. 


\section{Generalised dual descriptions of linearised gravity}             
In five dimensions the usual electromagnetic duality between the local descriptions in terms of the one and two-form gauge fields $A_{\mu}$ or $B_{\mu\nu}$ generalises to a gravitational {\textit{triality}} between the local descriptions in terms of the three gauge fields $h_{\mu\nu}$, $d_{\mu\nu\rho}$ or $c_{\mu\nu\rho\sigma}$ in representations corresponding to Young tableaux with two columns. The electromagnetic duality can be considered in arbitrary spacetime dimension $D$ with equivalent descriptions in terms of the electric or magnetic potentials $A_{\mu}$ or $B_{\mu_1 ... \mu_{D-3}}$. As discussed in {\cite{key5}}, one can similarly consider a linearised gravitational triality in $D$ dimensions with the conventional presentation in terms of the graviton $h_{\mu\nu}$ having equivalent descriptions in terms of either $D_{\mu_{1} ... \mu_{D-3} \nu}$ or $C_{\mu_{1} ... \mu_{D-3} \nu_1 ... \nu_{D-3}}$. The three dual fields correspond to the $GL(D,{\mathbb{R}})$-irreducible two-column Young tableaux representations $[1,1]$, $[D-3,1]$ and $[D-3,D-3]$.

The purpose of this section is to develop the theory of bi-forms which describe these gauge fields and their triality most economically. This work was first presented in {\cite{key6}}. 


\subsection{Bi-forms}
Consider the $GL(D,{\mathbb{R}})$-reducible tensor product space of $p$-forms and $q$-forms $X^{p,q} := \Lambda^p \otimes\Lambda^q$ on ${\mathbb{R}}^D$ whose elements are                    
%
%
\begin{equation} 
T \;\; =\;\; \frac{1}{p!q!}T_{\mu_1 ...\mu_p\nu_1 ...\nu_q} dx^{\mu_1}\wedge ...\wedge dx^{\mu_p}\otimes dx^{\nu_1}\wedge ...\wedge dx^{\nu_q}
\label{eq:2.1}
\end{equation} 
where the components $T_{\mu_1 ...\mu_p\nu_1 ...\nu_q} = T_{[ \mu_1 ...\mu_p ][ \nu_1 ...\nu_q ]}$ are totally antisymmetric in each of the $\{ \mu \}$ and $\{ \nu \}$ sets of indices separately. No symmetry properties are assumed between the indices $\mu _i$ and the indices $\nu_j$. The tensor field $T\in X^{p,q}$ is well defined and will be referred to as a {\textit{bi-form}}. This definition of a bi-form is useful since one can employ various constructions from the theory of forms acting on the individual $\Lambda^p$ and $\Lambda^q$ subspaces.  

A generalisation of the exterior wedge product defines the bi-form $T \mul T^{\,\prime} \in X^{p+ p^{\prime} ,q+ q^{\prime}}$, for any $T\in X^{p,q}$ and $T^{\,\prime} \in X^{ p^{\prime} , q^{\prime} }$, by
%
%
\begin{eqnarray}
T \mul T^{\,\prime} &=& \frac{1}{(p+ p^{\prime} )!(q+ q^{\prime} )!} 
T_{\mu_1 ...\mu_p\nu_1 ...\nu_q}{T^{\,\prime}}_{\mu_{p+1} ...\mu_{p+ p^{\prime}}\nu_{q+1} ...\nu_{q+
q^{\prime}}} dx^{\mu_1}\wedge ...\wedge dx^{\mu_{p+ p^{\prime}}} \nonumber \\ 
&&\hspace*{3.3in} \otimes \, dx^{\nu_1}\wedge ...\wedge dx^{\nu_{q+ q^{\prime}}}  
\label{eq:2.2}
\end{eqnarray}
This definition gives the space $X^* := \sum_{(p,q)} \oplus X^{p,q}$ a ring structure with respect to the $\mul$-product and the natural addition of bi-forms. 

Clearly, standard operations on differential forms generalise to bi-forms. There are two exterior derivatives on $X^{p,q}$. The left derivative
%
%
\be 
d : X^{p,q} \rightarrow X^{p+1,q}
\label{eq:2.3}
\ee
and the right derivative
%
\be
{\tilde{d}} : X^{p,q} \rightarrow X^{p,q+1}
\label{eq:2.4}
\ee
whose actions on $T$ are defined by
%
%
\begin{eqnarray}
dT &=& \frac{1}{p!q!}\partial_{[\mu}T_{\mu_1 ...\mu_p ]\nu_1 ...\nu_q} dx^{\mu}\wedge dx^{\mu_1}\wedge ...\wedge dx^{\mu_p}\otimes dx^{\nu_1}\wedge ...\wedge dx^{\nu_q} \, \in\, X^{p+1,q} \nonumber \\ 
{\tilde{d}} T &=& \frac{1}{p!q!}\partial_{[\nu}T_{|\mu_1 ...\mu_p |\nu_1 ...\nu_q ]} dx^{\mu_1}\wedge ...\wedge dx^{\mu_p}\otimes dx^{\nu}\wedge dx^{\nu_1}\wedge ...\wedge dx^{\nu_q} \, \in\, X^{p,q+1} 
\label{eq:2.5}
\end{eqnarray}
\footnote{As usual, the square bracketed indices are to be antisymmetrised while those inside vertical bars are excluded from the antisymmetrisation. For example $T_{[ \mu | \nu\rho | \sigma ]} := {\frac{1}{2}} ( T_{\mu\nu\rho\sigma} - T_{\sigma\nu\rho\mu} )$ for some fourth rank tensor $T_{\mu\nu\rho\sigma}$.} 
It is clear from these definitions that
%
%
\be
d^2 \; = \; {\tilde{d}}^2 \; = \; 0,\qquad d{\tilde{d}} \; = \; {\tilde{d}}d
\label{eq:2.6}
\ee
One can also write the total derivative
%
%
\be
{\cal{D}} : X^{p,q} \rightarrow X^{p+1,q} \oplus X^{p,q+1}
\label{eq:2.7}
\ee
defined as
%
%
\be
{\cal{D}} \; := \; d+ {\tilde{d}}
\label{eq:2.8}
\ee
which satisfies ${\cal{D}}^3 =0$.

Similarly, one can also construct distinct left 
%
%
\be
\iota_{k} : X^{p,q} \rightarrow X^{p-1,q}
\label{eq:2.9}
\ee
and right
%
%
\be
\tilde{\iota}_{k} : X^{p,q} \rightarrow X^{p,q-1}
\label{eq:2.10}
\ee
interior products defined by
%
%
\begin{eqnarray}
\iota_{k}T &=& \frac{1}{(p-1)!q!} k^{\mu_1} T_{\mu_1 \mu_2 ...\mu_p\nu_1 ...\nu_q} dx^{\mu_2}\wedge ...\wedge dx^{\mu_p}\otimes dx^{\nu_1}\wedge ...\wedge dx^{\nu_q} \, \in\, X^{p-1,q} \nonumber \\
\tilde{\iota}_{k}T &=& \frac{1}{p!(q-1)!} k^{\nu_1} T_{\mu_1 ...\mu_p\nu_1 \nu_2 ...\nu_q} dx^{\mu_1}\wedge ...\wedge dx^{\mu_p}\otimes dx^{\nu_2}\wedge ...\wedge dx^{\nu_q} \, \in\, X^{p,q-1} 
\label{eq:2.11}
\end{eqnarray}
for some vector field $k$. Again, it is clear that $\iota_{k}^2 = \tilde{\iota}_{k}^2 =0$ and $\iota_{k} \tilde{\iota}_{k}=\tilde{\iota}_{k} \iota_{k}$.

Consider now such bi-forms as reducible representations of the Lorentz group $SO(D-1,1) \subset GL(D,{\mathbb{R}})$, so that there is a Minkowski metric $\eta_{\mu\nu}$ and a totally antisymmetric tensor $\epsilon_{\mu_1 ... \mu_D}$ which are $SO(D-1,1)$-invariant tensors. These allow the construction of two inequivalent Hodge duality operations on bi-forms. There is a left dual
%
%
\be
*: X^{p,q} \rightarrow X^{D-p,q}
\label{eq:2.12}
\ee
and a right dual
%
%
\be
{\tilde{*}}: X^{p,q} \rightarrow X^{p,D-q}
\label{eq:2.13}
\ee
defined by
%
%
\begin{eqnarray} *T &=& \frac{1}{p!(D-p)!q!} T_{\mu_1 ...\mu_p\nu_1 ...\nu_q} 
\epsilon^{\mu_1 ...\mu_p}_{\;\;\;\;\;\;\;\;\;\;\mu_{p+1} ...\mu_D} dx^{\mu_{p+1}}\wedge ...\wedge
dx^{\mu_D}\otimes dx^{\nu_1}\wedge ...\wedge dx^{\nu_q} \, \in\, X^{D-p,q} \nonumber \\
\tilde{*}T &=& \frac{1}{p!q!(D-q)!} T_{\mu_1 ...\mu_p\nu_1 ...\nu_q} \epsilon^{\nu_1
...\nu_q}_{\;\;\;\;\;\;\;\;\;\;\nu_{q+1} ...\nu_D} dx^{\mu_1}\wedge ...\wedge dx^{\mu_p}\otimes
dx^{\nu_{q+1}}\wedge ...\wedge dx^{\nu_D} \, \in\, X^{p,D-q} \nonumber \\
\label{eq:2.14}
\end{eqnarray}
where indices are raised using the $SO(D-1,1)$-invariant metric. These definitions imply $*^2 = (-1)^{1+p(D-p)}$, ${\tilde{*}}^2 = (-1)^{1+q(D-q)}$ and $*\tilde{*}=\tilde{*}*$.
\footnote{Another formalism used to describe higher spin gauge theories is proposed in {\cite{key7}}, {\cite{key8}}, {\cite{key9}}. The construction there specifies a sequence of Young diagrams with increasing numbers of cells. There is then just one notion of exterior derivation which maps to the next element in the sequence and one interior product which maps to the previous element. This encounters some difficulties when it comes to describing the linearised gauge theory. Recall, for example, that the gauge transformation for $d_{\mu\nu\rho}$ in ({\ref{eq:1}}) contained terms involving two parameters $\alpha_{\mu\nu}$ and $\beta_{\mu\nu}$ with different index symmetries. Consequently if $d_{\mu\nu\rho}$ is an element in the complex then one could only write its gauge transformation as the exterior derivative of either the symmetric part of $\alpha_{\mu\nu}$ or $\beta_{\mu\nu}$ - whichever was chosen to be in the complex.}

This allows one to also define two inequivalent \lq adjoint' derivatives
%
%
\be
d^{\dagger} \; := \; {(-1)}^{1+D(p+1)} *d* : X^{p,q} \rightarrow X^{p-1,q} 
\label{eq:2.15}
\ee
and
%
%
\be 
{\tilde{d^{\dagger}}} \; := \; {(-1)}^{1+D(q+1)} {\tilde{*}} \, {\tilde{d}} {\tilde{*}} : X^{p,q} \rightarrow X^{p,q-1}
\label{eq:2.16}
\ee
whose actions on $T$ are defined by
%
%
\begin{eqnarray}
d^{\dagger} T &=& \frac{1}{(p-1)!q!}\partial^{\mu_1} T_{\mu_1 \mu_2 ... \mu_p \nu_1 ... \nu_q} dx^{\mu_2}\wedge ...\wedge dx^{\mu_p} \otimes dx^{\nu_1} \wedge ...\wedge dx^{\nu_q} \, \in\, X^{p-1,q} \nonumber \\ 
{\tilde{d^{\dagger}}} T &=& \frac{1}{p!(q-1)!} \partial^{\nu_1} T_{\mu_1 ... \mu_p \nu_1 \nu_2 ... \nu_q} dx^{\mu_1} \wedge ... \wedge dx^{\mu_p} \otimes dx^{\nu_2} \wedge ...\wedge dx^{\nu_q} \, \in\, X^{p,q-1} 
\label{eq:2.17}
\end{eqnarray}
These definitions imply ${d^\dagger}^2 = {\tilde{d^\dagger}}^2 =0$ and $d^\dagger {\tilde{d^\dagger}} = {\tilde{d^\dagger}} d^\dagger$. One can then define the Laplacian operator
%
%
\be
\Delta \; := \; d d^\dagger + d^\dagger d \; \equiv\; {\tilde{d}} {\tilde{d^\dagger}} + {\tilde{d^\dagger}} {\tilde{d}} : X^{p,q} \rightarrow X^{p,q}
\label{eq:E.1}
\ee
where the second equality follows identically.

A trace operation
%
%
\be
\tau : X^{p,q} \rightarrow X^{p-1,q-1}
\label{eq:2.18}
\ee
can be defined by
%
%
\begin{equation} 
\tau T \;\; =\;\; \frac{1}{(p-1)!(q-1)!} \eta^{\mu _1\nu_1} T_{\mu_1 ... \mu_p \nu_1 ... \nu_q} dx^{\mu_2} \wedge
... \wedge dx^{\mu_p} \otimes dx^{\nu_2} \wedge ... \wedge dx^{\nu_q} \, \in\, X^{p-1,q-1}
\label{eq:trace}
\end{equation}
Consequently, one can define two inequivalent \lq dual trace' operations
%
%
\be 
\sigma \; := \; {(-1)}^{1+D(p+1)} * \tau * : X^{p,q} \rightarrow X^{p+1,q-1}
\label{eq:2.20}
\ee
and
%
%
\be  
{\tilde{\sigma}} \; := \; {(-1)}^{1+D(q+1)} {\tilde{*}} \, \tau {\tilde{*}} : X^{p,q} \rightarrow X^{p-1,q+1}
\label{eq:2.21}
\ee
so that
%
%
\begin{eqnarray} 
\sigma T &=& \frac{{(-1)}^{p+1}}{p!(q-1)!} T_{[\mu_1 ...\mu_p \nu_1 ]\nu_2 ...\nu_q} dx^{\mu_1}\wedge ...\wedge
dx^{\mu_p}\wedge dx^{\nu_1} \otimes dx^{\nu_2}\wedge ...\wedge dx^{\nu_q} \, \in \, X^{p+1,q-1} \nonumber \\
{\tilde{\sigma}} T &=& \frac{{(-1)}^{q+1}}{(p-1)!q!}T_{\mu_1 ...\mu_{p-1}[\mu_p\nu_1 ...\nu_q]} dx^{\mu_1}\wedge
...\wedge dx^{\mu_{p-1}}\otimes dx^{\mu_p}\wedge dx^{\nu_1}\wedge ...\wedge dx^{\nu_q} \, \in \, X^{p-1,q+1} \nonumber \\
\label{eq:2.22}
\end{eqnarray} 

It is also useful to define a transpose operation
%
%
\be
t: X^{p,q} \rightarrow X^{q,p}
\label{eq:2.23}
\ee
by
%
%
\begin{equation} 
t T \;\; =\;\; \frac{1}{p!q!}T_{\nu_1 ...\nu_q \mu_1 ...\mu_p} dx^{\nu_1}\wedge ...\wedge dx^{\nu_q} \otimes dx^{\mu_1}\wedge ...\wedge dx^{\mu_p} \, \in \, X^{q,p}
\label{eq:2.24}
\end{equation}
and a map
%
%
\be 
\eta: X^{p,q} \rightarrow X^{p+1,q+1}
\label{eq:2.25}
\ee
by
%
%
\begin{equation} 
\eta T \;\; =\;\; \frac{1}{(p+1)!(q+1)!} \eta_{\mu_1 \nu_1} T_{ \mu_2 ...\mu_{p+1} \nu_2 ...\nu_{q+1}} dx^{\mu_1}\wedge ...\wedge dx^{\mu_p} \wedge dx^{\mu_{p+1}} \otimes dx^{\nu_1}\wedge ...\wedge dx^{\nu_q} \wedge dx^{\nu_{q+1}}
\label{eq:2.26}
\end{equation}
where the action in ({\ref{eq:2.26}}) is identical to the $\mul$-product with the $SO(D-1,1)$-invariant metric $\eta_{\mu\nu}$, so that $\eta T \equiv \eta \mul T$.

It is also convenient for the forthcoming discussion to state the identities
%
%
\begin{eqnarray}
\tau d + d \tau &=& {\tilde{d^\dagger}} \nonumber \\
\tau {\tilde{d}} + {\tilde{d}} \tau &=& d^\dagger \nonumber \\
\tau d^\dagger + d^\dagger \tau &=& 0 \label{eq:2.27} \\
\tau {\tilde{d^\dagger}} + {\tilde{d^\dagger}} \tau &=& 0 \nonumber 
\end{eqnarray}
which imply the further relations
%
%
\begin{eqnarray}
{(-1)}^{n+1} \tau^n d + d \tau^n &=& n\, {\tilde{d^\dagger}} \tau^{n-1} \nonumber \\
{(-1)}^{n+1} \tau^n {\tilde{d}} + {\tilde{d}} \tau^n &=& n\, d^\dagger \tau^{n-1} \nonumber \\
\sigma d + d \sigma &=& 0 \label{eq:2.28} \\
{\tilde{\sigma}} {\tilde{d}} + {\tilde{d}} {\tilde{\sigma}} &=& 0 \nonumber 
\end{eqnarray}
that follow by induction.

A further important result is that, for any bi-form $T \in X^{p,q}$, then
%
%
\begin{equation}
\tau^n T \;\; = \;\; 0 \;\;\;\; \Longrightarrow \;\;\;\; ( \tau^{D-p-q+n} *{\tilde{*}} ) T \;\; = \;\; 0 
\label{eq:2.29}
\end{equation}
but does not imply $( \tau^{D-p-q+n-1} *{\tilde{*}} ) T = 0$ for $n \geq 1$ and $D-p-q+n \geq 1$. The proof follows since the expression on the right hand side of ({\ref{eq:2.29}}) only contains terms with $n$ or more traces of $T$. Consequently $\tau T =0$ implies the whole bi-form $T =0$ for $D < p+q$ (but not for $D \geq p+q$). More generally, $\tau^n T =0$ implies $T=0$ for $D < p+q+1-n$.      
 

\subsection{Reducible gauge theories}

In this section, we will discuss the gauge theories for fields in reducible representations of $GL(D,{\mathbb{R}})$, and in the following section we will refine this to examine irreducible representations. Consider then a gauge field $A\in X^{p,q}$ with components $A_{\mu_1 ... \mu_p \nu_1 ... \nu_q} = A_{[ \mu_1 ... \mu_p ][ \nu_1 ... \nu_q ]}$. Consider also the general gauge transformation
%
%
\begin{equation}
\delta A \;\; = \;\; d \alpha^{p-1,q} + {\tilde{d}} {\tilde{\alpha}}^{p,q-1}  
\label{eq:3.1}
\end{equation}
with gauge parameters $\alpha^{p-1,q} \in X^{p-1,q}$, ${\tilde{\alpha}}^{p,q-1} \in X^{p,q-1}$. The gauge invariant field strength is
%
%
\be
F \; = \; d {\tilde{d}} A
\label{eq:3.2}
\ee
satisfying the Bianchi indentities
%
%
\be
dF \; = \; 0, \qq {\tilde{d}} F \; = \; 0
\label{eq:3.3}
\ee
A natural field equation to impose is
%
%
\be
\tau F \; = \; 0
\label{eq:3.4}
\ee
This gives a reducible theory. For example, $A\in X^{1,1}$ is a general second rank tensor $A_{\mu\nu}$ which can be decomposed into symmetric and antisymmetric parts. Such reducible theories were investigated in {\cite{key5}} and although the formalism can be developed, it seems more natural to decompose into irreducible representations of $GL(D,{\mathbb{R}})$.


\subsection{Bi-form gauge theory}
The formalism above can be used to describe a free gauge theory whose gauge potential $A\in X^{p,q}$ is a tensor field transforming in an {\textit{irreducible}} representation of $GL(D,{\mathbb{R}})$ such that its components $A_{ \mu_1 ... \mu_p \nu_1 ... \nu_q}$ have the index symmetry of a two-column Young tableau with $p$ cells in the left column and $q$ cells in the right column. Without loss of generality, we take $p \ge q$. A shorthand notation for this representation is $[p,q]$.
\footnote{The convention of writing tableaux in terms of the number of cells in each column differs from the standard way of labelling by row occupancy (e.g. in {\cite{key10}}), but is more suitable for the discussion here.}
We will write $X^{[p,q]}$ for the subspace of $X^{p,q}$ in the representation $[p,q]$. Recall that irreducibility under $GL(D,{\mathbb{R}})$ implies that the components of $A$ of must satisfy {\cite{key10}}
%
%
\begin{equation}
 \begin{array}{lrcl}
 & A_{[ \mu_1 ... \mu_p ][ \nu_1 ... \nu_q ]} &=& A_{\mu_1 ... \mu_p \nu_1 ... \nu_q} \;\;\;\;\;\; , \;\;\;\;\;\; A_{[ \mu_1 ... \mu_p \nu_1 ] \nu_2 ... \nu_q} \;\; =\;\; 0 \\
 {\mbox{and}}\;\;\;\;\;\;\;\; & A_{\mu_1 ... \mu_p \nu_1 ... \nu_q} &=& 
A_{\nu_1 ... \nu_q \mu_1 .. \mu_p} \;\;\;\;\;\;\;\; {\mbox{if}} \;\; p=q
 \end{array}
\label{eq:3.5}
\end{equation}
Such a tensor has
%
%
\begin{equation}
{dim}_D [p,q] \;\; =\;\; \left( \matrix{D \cr p \cr} \right) \left( \matrix{D+1 \cr q \cr} \right) \left[ 1- \frac{q}{p+1} \right]
\label{eq:3.6}
\end{equation}
independent components for $p \geq q$. For example, the graviton $h_{\mu\nu}$ in five dimensions has ${dim}_5 [1,1] = 15$ components. These conditions can be written as the conditions on the bi-form $A$
%
%
\begin{eqnarray}
\sigma A &=& 0 \nonumber \\
{\mbox{and}} \;\;\;\;\;\;\;\;\;\;\; A &=& tA \;\;\;\;\qq {\mbox{if}} \;\; p=q \label{eq:3.7}
\end{eqnarray}
Thus, for $p\ne q$, $X^{[p,q]}$ is the kernel of the map $\sigma :X^{p,q} \to X^{p+1,q-1}$, while for $p=q$ it is the subspace of the kernel invariant under the transpose $t$. It should be noted that ${\tilde{\sigma}} A \ne 0$ for $p >q$ though for $p=q$, $\sigma A = {\tilde{\sigma}} A=0$, since $tA =A$. 

This presentation is given before gauge fixing. After restricting to the physical (light-cone) gauge, the components of ${\hat{A}}$, written ${\hat{A}}_{i_1 ... i_p j_1 ... j_q}$, transform irreducibly under the little group $SO(D-2) \subset SO(D-1,1) \subset GL(D,{\mathbb{R}})$ (for $D \geq 2$). Since this implies the existence of an $SO(D-2)$-invariant metric $\delta_{ij}$ then irreducibility under $SO(D-2)$ implies that, in addition to the index symmetries ({\ref{eq:3.5}}), the components of ${\hat{A}}$ must also satisfy the tracelessness condition ${\hat{A}}^{i}_{\;\; i_2 ... i_p i j_2 ... j_q} =0$ (or equivalently $\tau {\hat{A}}=0$ with respect to the $SO(D-2)$-invariant metric). Such a representation, in physical gauge, therefore has
%
%
\begin{equation}
{dim}_{(D-2)} {\widehat{[p,q]}} \;\; = \;\; {dim}_{(D-2)} [p,q] - {dim}_{(D-2)} [p-1,q-1]
\label{eq:3.8}
\end{equation}
independent components. For example, the physical graviton ${\hat{h}}_{ij}$ in five dimensions has ${dim}_3 {\widehat{[1,1]}} = 5$.

The $GL(D,{\mathbb{R}})$-reducible space of bi-forms $X^{p,q}$ contains the space of all type $[p,q]$ tensors, written $X^{[p,q]}$, as a $GL(D,{\mathbb{R}})$-irreducible subspace so the bi-form operations in section 2.1 are well defined on these irreducible tensors. The projection from $X^{p,q}$ onto $X^{[p,q]}$ is the Young symmetriser ${\cal{Y}}_{[p,q]}$ for the particular $[p,q]$ tableau symmetry {\cite{key10}}. Thus $A$ satisfies
%
%
\be 
A \; = \; {\cal{Y}}_{[p,q]} \circ A
\label{eq:3.9}
\ee
In order to perserve this, we project the gauge transformation ({\ref{eq:3.1}}) to obtain
%
%
\begin{equation}
\delta A \;\; = \;\; {\cal{Y}}_{[p,q]} \circ \left( \, d \alpha^{p-1,q} + {\tilde{d}} {\tilde{\alpha}}^{p,q-1} \, \right)
\label{eq:3.10}
\end{equation}
for bi-form parameters $\alpha^{p-1,q} \in X^{p-1,q}$ and ${\tilde{\alpha}}^{p,q-1} \in X^{p,q-1}$. These gauge parameters are not assumed to be $GL(D,{\mathbb{R}})$-irreducible. The first order gauge transformation for $A$ is then proportional to the sum of two tableaux with symmetry of $[p,q]$ type - one term having a single partial derivative entered in the left column and the other having a single partial derivative entered in the right column. 

In conventional Abelian gauge theory with a $p$-form (one-column $[p,0]$ tableau) potential $A_{\mu_1 ... \mu_p}$ one can write a field strength $(p+1)$-form $F=dA$ which is invariant under the gauge transformation $\delta A = d \alpha$ for some $(p-1)$-form parameter $\alpha$. Following this, several authors {\cite{key1}}, {\cite{key11}}, {\cite{key12}} have proposed a type $[3,1]$ field strength involving a single derivative of the two-column, type $[2,1]$ gauge potential $d_{[\mu\nu ]\rho}$ used to describe the Pauli-Fierz system. Such a construction, however, is found to be invariant under only the $\alpha^{1,1}$ part of the most general gauge transformation proposed in ({\ref{eq:3.10}}). The example of the type $[1,1]$ graviton though has the linearised Riemann tensor, involving two derivatives, as its invariant field strength. This is indicative of the general observation that two-column tableaux gauge potentials should describe linearised \lq gravitational type' systems whose field strengths involve two derivatives of the given gauge potential. Consequently, the unique
$GL(D,{\mathbb{R}})$-irreducible field strength $F$ is the type $[p+1,q+1]$ tensor defined by
%
%
\begin{equation}
F \;\; = \;\; {\cal{Y}}_{[p+1,q+1]} \circ \left( \, d {\tilde{d}} A \, \right) \;\; \equiv \;\; d {\tilde{d}} A
\label{eq:3.11}
\end{equation}
which is invariant under the full gauge transformation ({\ref{eq:3.10}}). The left and right exterior derivatives act as $d : X^{[p,q]} \rightarrow X^{p+1,q}$ and ${\tilde{d}} : X^{[p,q]} \rightarrow X^{p,q+1}$ on the irreducible subspace $X^{[p,q]}$ and so do not map tableaux to tableaux. The composite operator $d \tilde{d}$, however, acts as $d{\tilde{d}} : X^{[p,q]} \rightarrow X^{[p+1,q+1]}$ on the irreducible subspace which implies the identity in ({\ref{eq:3.11}}). The expression is unambiguous since the left and right exterior derivatives commute. A theorem in {\cite{key8}}, in fact, allows any globally defined type $[p+1,q+1]$ tensor $F$ that is closed under both $d$ and $\tilde{d}$ to be written as in ({\ref{eq:3.11}}) for some locally defined type $[p,q]$ potential $A$ (this is true globally on ${\mathbb{R}}^D$). This generalises the well known Poincar\'{e} lemma that any closed $(p+1)$-form can be written locally as the exterior derivative of some $p$-form potential. In terms of Young tableaux, ({\ref{eq:3.11}}) simply corresponds to a type $[p+1,q+1]$ pattern with one partial derivative in each of the two columns. The gauge invariance of ({\ref{eq:3.11}}) then follows because $\delta F$ corresponds to a type $[p+1,q+1]$ pattern with at least two (commuting) partial derivatives in a single column.

In addition to this gauge invariance, the field strength $F$ also satisfies the two second Bianchi identities
%
%
\begin{equation}
dF \; =\; 0, \qq \;\;\; {\tilde{d}} F \; =\; 0 
\label{eq:3.12}
\end{equation}
which follow from a similar reasoning, and the first Bianchi identity
%
%
\be 
\sigma F \; = \; 0
\label{eq:3.13}
\ee
for $p \geq q$, by virtue of $GL(D,{\mathbb{R}})$-irreducibility.

The natural equation of motion is the generalised Einstein equation
%
%
\be
\tau F\; =\; 0
\label{eq:3.14}
\ee
This is non-trivial, in the sense that it does not imply $F=0$, for dimension $D\geq p+q+2$ (using ({\ref{eq:2.29}})). More generally, for dimensions $D = p+q+3-n$, the field equation
%
%
\be
\tau ^n F \; =\; 0
\label{eq:3.15}
\ee
gives a non-trivial equation. For example, for a type [1,1] tensor (graviton) $h_{\mu\nu}$, $F$ is the type [2,2] linearised Riemann tensor, and $\tau F=0$ is the vacuum Einstein equation $R_{\mu\nu}=0$. This is non-trivial for $D \geq 4$. For $D=3$, the Einstein equation $R_{\mu\nu}=0$ implies the whole curvature vanishes, and is too strong as it has no non-trivial solutions. For $D=3$, the equation $\tau^2F=0$, i.e. the vanishing of the Ricci scalar $R=0$ does give a non-trivial equation and gives a theory dual to a scalar field {\cite{key5}}.


\subsection{Dualities between type $[p,q]$ tensors}
As discussed in {\cite{key5}}, although the gravitational triality in five dimensions was found via dimensional reduction of a self-dual field in six dimensions, one can more generally consider the dual descriptions of a graviton in $D$ dimensions. Given the linearised Riemann curvature $R$ as the type $[2,2]$ field strength ({\ref{eq:3.11}}) for the graviton then one can construct two inequivalent Hodge duals given by the bi-forms $S := *R \, \in \, X^{D-2,2}$ and $G := *{\tilde{*}} R \, \in \, X^{D-2,D-2}$ which are written
%
%
\begin{eqnarray} S_{\mu_1 ... \mu_{D-2} \nu_1 \nu_{2}} &=& \frac{1}{2} \, R^{\alpha\beta}_{\;\;\;\;\; \nu_1
\nu_2} \, \epsilon_{\alpha\beta \mu_1 ... \mu_{D-2}} \label{eq:3.16} \\ G_{\mu_1 ... \mu_{D-2} \nu_1 ...
\nu_{D-2}} &=& \frac{1}{4} \, R^{\alpha\beta\gamma\delta} \, \epsilon_{\alpha\beta \mu_1 ... \mu_{D-2}} \,
\epsilon_{\gamma\delta \nu_1 ... \nu_{D-2}} \label{eq:3.17}
\end{eqnarray}
in component form.

The curvature tensor satisfies the first Bianchi identities
%
%
\be
\sigma R\; =\; 0, \qq\qq {\tilde{\sigma}}R\; =\; 0
\label{eq:3.18}
\ee
the second Bianchi identities
%
%
\be
dR\; =\; 0,\qq\qq {\tilde{d}}R\; =\; 0
\label{eq:3.19}
\ee
and the Einstein equation
%
%
\be
\tau R\; =\; 0
\label{eq:3.20}
\ee
in $D \geq 4$, which implies the secondary field equations
%
%
\be
d^\dagger R\; =\; 0,\qq\qq {\tilde{d^\dagger}} R\; =\; 0
\label{eq:3.21}
\ee
which are obtained using ({\ref{eq:2.27}}).

The dual tensors
%
%
\be
S\; =\; *R, \qq\qq G\; =\; *{\tilde{*}}R
\label{eq:3.22}
\ee
then satisfy the algebraic constraints
%
%
\be
\sigma S \; =\; 0, \qq\qq \sigma G\; =\; {\tilde{\sigma}}G\; =\; 0
\label{eq:3.23}
\ee
which imply they are $GL(D,{\mathbb{R}})$-irreducible, such that $S \in X^{[D-2,2]}$ and $G \in X^{[D-2,D-2]}$. They also satisfy the differential constraints
%
%
\be
dS \; =\; 0, \qq {\tilde{d}}S\; =\; 0, \qq d^\dagger S\; =\; 0, \qq {\tilde{d^\dagger}}S\; =\; 0
\label{eq:3.24}
\ee
and 
%
%
\be
dG \; =\; 0, \qq {\tilde{d}}G\; =\; 0, \qq \tau^{D-4} d^\dagger G\; =\; 0, \qq \tau^{D-4} {\tilde{d^\dagger}}G\; =\; 0
\label{eq:3.25}
\ee
together with the field equations
%
%
\be
\tau S \; =\; 0, \qq\qq \tau^{D-3} G\; =\; 0
\label{eq:3.26}
\ee
which are non-trivial for $S$ and $G$ in $D \geq 4$. These are derived from the properties of the Riemann tensor and by using ({\ref{eq:2.28}}) and ({\ref{eq:2.29}}). The dualities ({\ref{eq:3.16}}) and ({\ref{eq:3.17}}) interchange field equations and Bianchi identities. For example the field equation $\tau R=0$ becomes $\tau *S=0$ and which is equivalent to $*\sigma S=0$, implying the Bianchi identity $\sigma S=0$. It then gives the field equation $\tau *{\tilde{*}} G=0$, which is equivalent to $\tau^{D-3} G=0$. The Bianchi identity $\sigma R=0$ becomes the field equation $\tau S=0$ and then the Bianchi identity ${\tilde{\sigma}} G=0$. These properties follow from the fact that $\tau$, which occurs in the field equations, is related to $\sigma$, which occurs in the Bianchi identities, by Hodge duality as $\sigma \sim *\tau *$.
 
The irreducibility of the duals $S$ and $G$, together with the fact that they are closed under both $d$ and ${\tilde{d}}$ further implies that these tensors can be solved in terms of the type $[D-3,1]$ and $[D-3,D-3]$ gauge potentials $D_{\mu_1 ... \mu_{D-3} \nu}$ and $C_{\mu_1 ... \mu_{D-3} \nu_1 ... \nu_{D-3}}$, such that $S = d {\tilde{d}} D$ and $G= d {\tilde{d}} C$. The three constraints $\tau R =0$, $\tau S=0$ and $\tau^{D-3} G=0$ can then be seen as the non-trivial, linearised equations of motion for $h$, $D$ and $C$.

Another way to see the duality given above is in physical gauge where one has no gauge symmetry to consider and there exists an $SO(D-2)$ orientation tensor $\epsilon^{i_1 ... i_{D-2}}$. The two dual potentials are then related to the physical graviton $h_{ij}$ such that $D=*h$ and $C=*{\tilde{*}}h$ (where $*$ here denotes the $SO(D-2)$-covariant Hodge dual). This gauge-fixed definition of $D$ and $C$ implies that they are irreducible under $SO(D-2)$ following the irreducibility of $h$. A non-trivial check that each of these three fields describes the same number of physical degrees of freedom follows the result that
%
%
\be
{dim}_{(D-2)} {\widehat{[1,1]}} \;\; = \;\; {dim}_{(D-2)} {\widehat{[D-3,1]}} \;\; = \;\; {dim}_{(D-2)} {\widehat{[D-3,D-3]}} \;\; = \;\; \frac{D(D-3)}{2}
\label{eq:3.27}
\ee
for $D > 3$.
\footnote{This is analogous to the more usual duality in physical gauge wherein a $p$-form gauge field $A$ is related to a dual $(D-2-p)$-form $B$ such that $B=*A$. In this case both fields describe the same $\left( \matrix{D-2 \cr p \cr} \right)$ physical degrees of freedom.}

The arguments presented in this section are straightforwardly extended for dualities between arbitrary type $[p,q]$ tensor gauge theories. Consider a type $[p,q]$ tensor gauge field $h$ (with $p \geq q$) having type $[p+1,q+1]$ field strength $R = d{\tilde{d}} h$ satisfying the first and second Bianchi identities $\sigma R=0$, $dR =0$ and ${\tilde{d}} R=0$. It is assumed that $D \geq p+q+2$ so that the \lq Einstein' equation $\tau R=0$ which is imposed has non-trivial solutions. These properties of $R$ again imply the secondary field equations $d^\dagger R =0$ and ${\tilde{d^\dagger}} R=0$.

For general $p \geq q$ one can define three inequivalent Hodge dual field strengths
%
%
\be
S\; :=\; *R \, \in \, X^{D-p-1,q+1}, \qq {\tilde{S}}\; :=\; {\tilde{*}}R \, \in \, X^{p+1,D-q-1}, \qq G\; :=\; *{\tilde{*}} R \, \in \, X^{D-p-1,D-q-1} 
\label{eq:3.28}
\ee
where ${\tilde{S}} = tS$ for $p=q$. The above dualities, together with the properties of $R$, imply the algebraic constraints
%
%
\be
\sigma S \; =\; 0, \qq\qq {\tilde{\sigma}} {\tilde{S}}\; =\; 0, \qq\qq {\tilde{\sigma}} G\; =\; 0
\label{eq:3.29}
\ee
with the additional constraints $\sigma G=0$ only if $p=q$ and ${\tilde{\sigma}} S = \sigma {\tilde{S}} =0$ only if $D=p+q+2$. $S$, ${\tilde{S}}$ and $G$ are therefore $GL(D,{\mathbb{R}})$-irreducible.
\footnote{The constraints in ({\ref{eq:3.29}}) imply this following the assumption $D \geq p+q+2$ which implies that the bi-form $S$ has a left column length $\geq$ that of the right column and vice versa for ${\tilde{S}}$. The bi-form $G$ has right column length $\geq$ left column length, due to the assumption $p \geq q$.}
The dual tensors also satisfy the differential constraints
%
%
\begin{equation}
 \begin{array}{rcl}
 dS &=& 0, \qq\qq d^\dagger S \;\; =\;\; 0, \qq\qq {\tilde{d^\dagger}}S \;\; =\;\; 0, \\
 d{\tilde{S}} &=& 0, \qq\qq\; {\tilde{d}}{\tilde{S}} \;\; =\;\; 0, \qq\qq {\tilde{d^\dagger}}{\tilde{S}} \;\; =\;\; 0, \\
 {\tilde{d}}G &=& 0, \qq \tau^{D-2-p-q} d^\dagger G \;\; =\;\; 0 
 \end{array}
\label{eq:3.30}
\end{equation}
with the additional constraints ${\tilde{d}}S =0$, $d^\dagger {\tilde{S}}=0$, $dG=0$ and $\tau^{D-2-p-q} {\tilde{d^\dagger}} G=0$ if $p=q$. The field equations imposed on the dual tensors are
%
%
\be
\tau S\; =\; 0, \qq\qq \tau^{1+p-q} {\tilde{S}}\; =\; 0, \qq\qq \tau^{D-1-p-q} G\; =\; 0 
\label{eq:3.31}
\ee
which are non-trivial equations for $S$, ${\tilde{S}}$ and $G$ if $p \geq q$. The expressions ({\ref{eq:2.28}}) and ({\ref{eq:2.29}}) are used to derive the results above and again the dualities mix Bianchi and field equation constraints for the dual field strengths. For $p=q$, the expressions above imply $S$($=t{\tilde{S}}$) and $G$ can be solved in terms of type $[D-p-2,p]$ and $[D-p-2,D-p-2]$ tensor potentials $D$ and $C$, such that $S = d{\tilde{d}}D$ and $G=d{\tilde{d}}C$. The field equations ({\ref{eq:3.31}}) can then be seen as the equations of motion for $D$ and $C$. 


\subsection{Gauge invariant actions}
In this section we construct the gauge invariant actions corresponding to the field equations given above for bi-form gauge
fields in sufficiently large dimension. The construction of such actions is perhaps best described by starting with some simple examples. Consider the three fields $h_{\mu\nu}$, $d_{\mu\nu\rho}$ and $c_{\mu\nu\rho\sigma}$ of types $[1,1]$, $[2,1]$ and $[2,2]$ described in section 1, but now in $D$ dimensions.

The invariant action ${\cal{S}}^{[1,1]}$ for the graviton $h_{\mu\nu}$ is the linearised Einstein-Hilbert action given by
%
%
\begin{equation}
{\cal{S}}^{[1,1]} \;\; = \;\; - \frac{1}{2} \int d^D x \, h^{\mu\nu} E_{\mu\nu} \;\; = \;\; \int d^D x \, \left( \, \partial^{[ \mu} h^{\nu ] \rho} \partial_{[ \mu} h_{\nu ] \rho} -2 \partial^{[ \mu} h^{\nu ] \nu} \partial_{[ \mu} h_{\rho ] \rho} \, \right)  
\label{eq:3.32}
\end{equation}
where $E_{\mu\nu} := R_{\mu\nu} - \frac{1}{2} \eta_{\mu\nu} R$ denotes the linearised Einstein tensor, satisfying $\partial^{\mu} E_{\mu\nu} \equiv 0$, constructed from contractions of the linearised Riemann tensor $R_{\mu\nu\rho\sigma} = -4
\partial_{[ \mu} h_{\nu ][ \rho , \sigma ]}$. The Lagrangian has two expressions as either $- \frac{1}{2} h^{\mu\nu} E_{\mu\nu}$ in terms of the linearised Einstein tensor $E_{\mu\nu}$ or as quadratic terms in the type $[2,1]$ single derivative object $\partial_{[\mu} h_{\nu ] \rho}$ and its $SO(D-1,1)$-irreducible trace part $\partial_{[\mu} h_{\nu ]}^{\;\;\;\; \nu}$. Notice that neither $\partial_{[\mu} h_{\nu ] \rho}$ nor its $SO(D-1,1)$-irreducible traceless part is invariant under $\delta h_{\mu\nu} = 2 \partial_{(\mu} \xi_{\nu )}$ even though ${\cal{S}}^{[1,1]}$ is. The easiest way to see the gauge invariance is in terms of the $- \frac{1}{2} h^{\mu\nu} E_{\mu\nu}$ Lagrangian. Since the linearised Riemann tensor $R_{\mu\nu\rho\sigma} = -4 \partial_{[\mu} h_{\nu ][ \rho ,\sigma ]}$ is the field strength ({\ref{eq:3.11}}) for the graviton then it follows that the linearised Einstein tensor is also gauge invariant (it being constructed from traces of $R_{\mu\nu\rho\sigma}$). The gauge transformation therefore only changes the Lagrangian by $- \frac{1}{2} ( \delta h_{\mu\nu} ) E^{\mu\nu}$ which simply consists of a total derivative term (since $\partial^{\mu} E_{\mu\nu} \equiv 0$ identically) and so vanishes in the action integral. The field equation for this model is simply the linearised vacuum Einstein equation $E_{\mu\nu} =0$ which is equivalent to the linearised Ricci flatness condition $R^{\alpha}_{\;\;\; \mu\alpha\nu} =0$ in $D \not= 2$. It is also noted that this equation is trivial unless $D \geq 4$.
\footnote{In the lower dimension $D=3$, one cannot obtain the non-trivial (vanishing Ricci scalar) equation $R=0$ from a gauge invariant action.}
The equation can be decomposed in terms of its linearly independent components, whence the graviton satisfies $\partial^2 h_{\mu\nu} =0$, $\partial^{\mu} h_{\mu\nu} =0$ and $h_{\mu}^{\;\; \mu} =0$.   
 
The gauge field $d_{\mu\nu\rho}$ has a field strength $S_{\mu\nu\rho\alpha\beta} = -6 \, \partial_{[ \mu} d_{\nu\rho ][ \alpha ,\beta ]}$ ({\ref{eq:3.11}}) and field equation $\tau S=0$. The invariant action ${\cal{S}}^{[2,1]}$ giving this field equation has been constructed in {\cite{key11}}, {\cite{key12}}. The presentation in these articles consists of quadratic terms in the type $[3,1]$ single derivative object $\partial_{[ \mu} d_{\nu\rho ] \alpha}$ and its trace part $\partial_{[ \mu} d_{\nu\alpha ]}^{\;\;\;\;\;\; \alpha}$. As already noted, these objects are not individually gauge invariant under ({\ref{eq:3.10}}), even though ${\cal{S}}^{[2,1]}$ is. A more obviously gauge invariant presentation is given by
%
%
\begin{equation}
{\cal{S}}^{[2,1]} \;\; =\;\; - \frac{1}{2} \int d^D x \; d^{\mu\nu\rho} \, E_{\mu\nu\rho}  
\label{eq:3.33}
\end{equation}
where $E_{\mu\nu\rho}$ is the linearised \lq Einstein' tensor for $d_{\mu\nu\rho}$, defined by 
%
%
\begin{equation}
E_{\mu\nu\rho} \;\; :=\;\; S^{\alpha}_{\;\;\; \mu\nu \alpha\rho} - \eta_{\rho [ \mu} \, S^{\alpha\beta}_{\;\;\;\;\; \nu ] \alpha\beta}   
\label{eq:3.34}
\end{equation}
or equivalently, $E= \tau S- \eta \tau^2 S$ in bi-form notation. By construction, $E_{\mu\nu\rho}$ is a gauge invariant type $[2,1]$ tensor satisfying $\partial^{\mu} E_{\mu\nu\rho} \equiv 0$ and $\partial^{\rho} E_{\mu\nu\rho} \equiv 0$ identically. Consequently the gauge transformation of the Lagrangian in ({\ref{eq:3.33}}) is a total derivative implying the action is gauge invariant. The field equation for this model is then the linearised vacuum \lq Einstein' equation
$E_{\mu\nu\rho} =0$ which is equivalent to $S^{\alpha}_{\;\;\; \mu\nu \alpha\rho} =0$ in $D \ne 3$. This equation is non-trivial in $D \geq 5$. The linearly independent components of this imply that $d_{\mu\nu\rho}$ satisfies $\partial^2 d_{\mu\nu\rho} =0$, $\partial^\mu d_{\mu\nu\rho} =0$, $\partial^\rho d_{\mu\nu\rho} =0$ and $d^{\;\;\;\;\; \mu}_{\mu\nu} =0$.

The invariant action ${\cal{S}}^{[2,2]}$ for $c_{\mu\nu\rho\sigma}$ also has a more explicitly gauge invariant form given by
%
%
\begin{equation}
{\cal{S}}^{[2,2]} \;\; =\;\; - \frac{1}{4} \int d^D x \; c^{\mu\nu\rho\sigma} \, E_{\mu\nu\rho\sigma}  
\label{eq:3.35}
\end{equation}
where $E_{\mu\nu\rho\sigma}$ is defined by
%
%
\begin{equation}
E_{\mu\nu\rho\sigma} \;\; :=\;\; G^{\alpha}_{\;\; \mu\nu \alpha\rho\sigma} - \eta_{\rho [ \mu} \, G^{\alpha\beta}_{\;\;\;\;\; \nu ] \alpha\beta \sigma} + \eta_{\sigma [ \mu} \, G^{\alpha\beta}_{\;\;\;\;\; \nu ] \alpha\beta \rho} + \frac{1}{3} \,  \eta_{\rho [ \mu} \, \eta_{\nu ] \sigma} \, G^{\alpha\beta\gamma}_{\;\;\;\;\;\; \alpha\beta\gamma}  
\label{eq:3.36}
\end{equation}
in terms of the field strength $G_{\mu\nu\rho\sigma\alpha\beta} = 9 \, \partial_{[ \mu} c_{\nu\rho ][ \sigma\alpha ,\beta ]}$. ({\ref{eq:3.36}}) can equivalently be written as $E=\tau G-2\eta \tau^2 G+{\frac{1}{3}} \eta^2 \tau^3 G$ in bi-form notation. By construction, $E_{\mu\nu\rho\sigma}$ is a gauge invariant type $[2,2]$ tensor satisfying $\partial^{\mu} E_{\mu\nu\rho\sigma} \equiv 0$. This implies the gauge invariance of ({\ref{eq:3.35}}). The field equation for this model is $E_{\mu\nu\rho\sigma} =0$ which is equivalent to $G^{\alpha}_{\;\; \mu\nu \alpha\rho\sigma} =0$ in $D \ne 4$. This equation is non-trivial in $D \geq 6$. The linearly independent components of this then imply that $c_{\mu\nu\rho\sigma}$ satisfies
$\partial^2 c_{\mu\nu\rho \sigma} =0$, $\partial^{\mu} c_{\mu\nu\rho\sigma} =0$ and $c_{\mu\nu \;\;\;\sigma}^{\;\;\;\;\; \mu} =0$.

The general construction then considers the gauge invariant action ${\cal{S}}^{[p,q]}$ for a given type $[p,q]$ tensor gauge field $A_{ \mu_1 ... \mu_p \nu_1 ... \nu_q}$. The gauge invariance can be seen by writing ${\cal{S}}^{[p,q]}$ in the form 
%
%
\begin{equation}
{\cal{S}}^{[p,q]} \;\; =\;\; - \frac{1}{p!q!} \int d^D x \; A^{ \mu_1 ... \mu_p \nu_1 ... \nu_q} \, E_{ \mu_1 ... \mu_p \nu_1 ... \nu_q}  
\label{eq:3.37}
\end{equation}
where $E_{ \mu_1 ... \mu_p \nu_1 ... \nu_q}$ is the gauge invariant type $[p,q]$ tensor satisfying $\partial^{\mu_1} E_{ \mu_1 ... \mu_p \nu_1 ... \nu_q} \equiv 0$ and $\partial^{\nu_1} E_{ \mu_1 ... \mu_p \nu_1 ... \nu_q} \equiv 0$ identically. Such a tensor always exists and involves various terms involving two derivatives on $A_{ \mu_1 ... \mu_p \nu_1 ... \nu_q}$ and all its possible traces. The general form of $E_{ \mu_1 ... \mu_p \nu_1 ... \nu_q}$ is given by 
%
%
\begin{eqnarray}
E_{\mu_1 ... \mu_p \nu_1 ... \nu_q} &=& {\cal{Y}}_{[p,q]} \circ \left( \, F^{\alpha}_{\;\;\; \mu_1 ... \mu_p \alpha \nu_1 ... \nu_q} + a_1 \, \eta_{\mu_1 \nu_1} F^{\alpha_1 \alpha_2}_{\;\;\;\;\;\;\;\;\; \mu_2 ... \mu_p \alpha_1 \alpha_2 \nu_2 ... \nu_q} \right. \label{eq:3.38} \\
&&\hspace*{.6in} \left. + \; .\, .\, . \; + a_{q} \, \eta_{\mu_1 \nu_1} ... \eta_{\mu_{q} \nu_{q}} F^{\alpha_1 ... \alpha_{q+1}}_{\;\;\;\;\;\;\;\;\;\;\;\;\;\;\;\; \mu_{q+1} ... \mu_p \alpha_1 ... \alpha_{q+1}} \, \right) \nonumber
\end{eqnarray}
for $q$ coefficients $a_1$,...,$a_{q}$ (assuming $p \geq q$). The leading order term is the single trace $\tau F$ of the type $[p+1,q+1]$ field strength $F$ of $A$ which is a type $[p,q]$ tensor. The correction terms involve successive traces of this object (appropriately symmetrised using ${\cal{Y}}_{[p,q]}$). The precise values of the coefficients are fixed uniquely by the conservation conditions above (for example, $a_1 = -pq/2$). The field equation is given by the linearised vacuum \lq Einstein' equation $E_{\mu_1 ... \mu_p \nu_1 ... \nu_q} =0$ which is equivalent to the equation $F^{\alpha}_{\;\;\; \mu_1 ... \mu_p \alpha \nu_1 ... \nu_q} =0$, proposed in {\cite{key5}}, for $D \not= p+q$. This equation is non-trivial in $D \geq p+q+2$. The linearly independent components of this equation imply $\partial^2 A_{ \mu_1 ... \mu_p \nu_1 ... \nu_q} =0$, $\partial^{\mu_1} A_{ \mu_1 ... \mu_p \nu_1 ... \nu_q} =0$, $\partial^{\nu_1} A_{ \mu_1 ... \mu_p \nu_1 ... \nu_q} =0$ and $A^{\alpha}_{\;\;\; \mu_2 ... \mu_p \alpha \nu_2 ... \nu_q} =0$.   


\section{Multi-form structure in exotic tensor gauge theory}
Having found the structure of bi-forms most suitable to describe gravitational dualities, this section will develop the general structure of multi-forms which can be used to construct gauge theories in $D$ dimensions whose gauge
fields transform in general irreducible representations of $GL(D,{\mathbb{R}})$. The various dual descriptions of such gauge theories are also considered. This generalised description could also be relevant in studies of string field theory and ${\cal{W}}$-geometry {\cite{key13}}. The multi-form construction has also been considered in {\cite{key9}}, {\cite{key14}}.


\subsection{Multi-forms}
A {\textit{multi-form}} of order $N$ is a tensor field $T$ that is an element of the $GL(D,{\mathbb{R}})$-reducible $N$-fold tensor product space of $p_i$-forms (where $i=1,...,N$), written 
%
%
\be
X^{p_1 ,..., p_N} \;\; := \;\; \Lambda^{p_1} \otimes ... \otimes \Lambda^{p_N}
\label{eq:4.1}
\ee
The components of $T$ are written $T_{{\mu}^1_1 ...{\mu}^1_{p_1} ...{\mu}^i_1 ...{\mu}^i_{p_i} ... {\mu}^N_{p_N} }$ and are taken to be totally antisymmetric in each set of $\{ {\mu}^i \}$ indices, such that
%
%
\be
T_{[ {\mu}^1_1 ...{\mu}^1_{p_1} ] ...[ {\mu}^i_1 ...{\mu}^i_{p_i} ]... [ {\mu}^N_{1} ... {\mu}^N_{p_N} ]} \;\; = \;\; T_{{\mu}^1_1 ...{\mu}^1_{p_1} ...{\mu}^i_1 ...{\mu}^i_{p_i} ... {\mu}^N_{1} ... {\mu}^N_{p_N} }
\label{eq:4.2}
\ee

The generalisation of the operations defined for bi-forms to multi-forms of order $N$ over ${\mathbb{R}}^D$ is then straightforward. The $\mul$-product is the map
%
%
\be
\mul : X^{p_1 ,...,p_N } \times X^{{p'}_1 ,...,{p'}_N } \rightarrow X^{p_1 +{p'}_1 ,...,p_N +{p'}_N }
\label{eq:4.3}
\ee
defined by the $N$-fold wedge product on the individual form subspaces, by analogy with ({\ref{eq:2.2}}).

There are $N$ inequivalent exterior derivatives 
%
%
\be
d^{(i)} : X^{p_1 ,...,p_i ,...p_N } \rightarrow X^{p_1 ,...,p_i +1 ,...,p_N }
\label{eq:4.4}
\ee
which are individually defined, by analogy with ({\ref{eq:2.5}}), as the exterior derivatives acting on the
$\Lambda^{p_i}$ form subspaces. This definition implies ${d^{(i)}}^2 =0$ (with no sum over $i$) and that any
two $d^{(i)}$ commute. One can also define the total derivative
%
%
\be
{\cal{D}} \; := \; \sum_{i=1}^{N} d^{(i)} : X^{p_1 ,...,p_i ,...p_N } \rightarrow \sum_{i=1}^{N} \oplus\, X^{p_1 ,...,p_i +1 ,...,p_N }
\label{eq:4.5}
\ee
which satisfies ${\cal{D}}^{N+1} =0$.

There are $N$ inequivalent interior products 
%
%
\be
{\iota}^{(i)}_{k} : X^{p_1 ,...,p_i ,...,p_N } \rightarrow X^{p_1 ,...,p_i -1 ,...,p_N }
\label{eq:4.6}
\ee
whose action is defined, by analogy with ({\ref{eq:2.11}}), as the individual interior products on each $\Lambda^{p_i}$ form subspace. Consequently ${{\iota}^{(i)}_{k}}^2 =0$ (with no sum over $i$) and any two ${\iota}^{(i)}_{k}$ commute.

For representations of $SO(D-1,1) \subset GL(D,{\mathbb{R}})$ there are $N$ inequivalent Hodge dual operations
%
%
\be
*^{(i)} : X^{p_1 ,...,p_i ,...p_N } \to X^{p_1 ,...,D- p_i ,...p_N }
\label{eq:4.7}
\ee
which, following ({\ref{eq:2.14}}), are defined to act as the Hodge duals on the individual $\Lambda^{p_i}$ form subspaces. This implies that ${*^{(i)}}^2 = (-1)^{1+ p_i (D-p_i )}$ (with no sum over $i$) with any two $*^{(i)}$  commuting.

This also allows one to define $N$ inequivalent \lq adjoint' exterior derivatives
%
%
\be
{d^\dagger}^{(i)} \; :=\; {(-1)}^{1+D( p_i +1)} *^{(i)} d^{(i)} *^{(i)} : X^{p_1 ,...,p_i ,...p_N } \rightarrow X^{p_1 ,...,p_i -1 ,...,p_N }
\label{eq:4.8}
\ee
This implies ${{d^\dagger}^{(i)}}^2 =0$ (with no sum over $i$) and any two ${d^\dagger}^{(i)}$ commute. One can then define the Laplacian operator
%
%
\be
\Delta \; :=\; d^{(i)} {d^\dagger}^{(i)} + {d^\dagger}^{(i)} d^{(i)} : X^{p_1 ,...,p_i ,...p_N } \rightarrow X^{p_1 ,...,p_i ,...,p_N }
\label{eq:4.9}
\ee
with no sum over $i$.

There exist $(N-1)!$ inequivalent trace operations
%
%
\be
\tau^{(ij)} : X^{p_1 ,...,p_i ,..., p_j ,...p_N } \rightarrow X^{p_1 ,...,p_i -1,..., p_j -1,...,p_N }
\label{eq:4.10}
\ee
defined, by analogy with ({\ref{eq:trace}}), as the single trace between the $\Lambda^{p_i}$ and $\Lambda^{p_j}$ form subspaces using $\eta^{{\mu}^i_1 {\mu}^j_1}$. This allows one to define two inequivalent \lq dual-trace' operations
%
%
\be
\sigma^{(ij)} \; :=\; {(-1)}^{1+D( p_i +1)} *^{(i)} \tau^{(ij)} *^{(i)} : X^{p_1 ,...,p_i ,..., p_j ,...p_N } \rightarrow X^{p_1 ,...,p_i +1,..., p_j -1,...,p_N }
\label{eq:4.11}
\ee
and
%
%
\be
{\tilde{\sigma}^{(ij)}} \; :=\; {(-1)}^{1+D( p_j +1)} *^{(j)} \tau^{(ij)} *^{(j)} : X^{p_1 ,...,p_i ,..., p_j ,...p_N } \rightarrow X^{p_1 ,...,p_i -1,..., p_j +1,...,p_N }
\label{eq:4.12}
\ee
associated with a given $\tau^{(ij)}$ (with no sum over $i$ or $j$). Notice that ${\tilde{\sigma}}^{(ij)} = \sigma^{(ji)}$ since $\tau^{(ij)} = \tau^{(ji)}$.

One can also write $(N-1)!$ inequivalent involutions
%
%
\be
t^{(ij)} : X^{p_1 ,...,p_i ,..., p_j ,...p_N } \rightarrow X^{p_1 ,...,p_j ,..., p_i ,...,p_N }
\label{eq:4.13}
\ee
defined by exchange of the $\Lambda^{p_i}$ and $\Lambda^{p_j}$ form subspaces in the tensor product space. And finally, there are $(N-1)!$ distinct operations
%
%
\be
\eta_{(ij)} : X^{p_1 ,...,p_i ,..., p_j ,...p_N } \rightarrow X^{p_1 ,...,p_i +1,..., p_j +1,...,p_N }
\label{eq:4.14}
\ee
defined as the $\mul$-product with the $SO(D-1,1)$ metric $\eta$ (understood as a $[1,1]$ bi-form in the $\Lambda^{p_i} \otimes \Lambda^{p_j}$ subspace), such that $\eta_{(ij)} T \equiv \eta \mul T$ for any $T \in X^{p_1 ,..., p_N}$. 


\subsection{Multi-form gauge theory}
Consider now a gauge potential that is a tensor $A$ in an arbitrary irreducible representation of $GL(D,{\mathbb{R}})$ whose components have the index symmetry of an $N$-column Young tableaux with $p_i$ cells in the $i$th column (it is assumed $p_i \geq p_{i+1}$). A given multi-form $A \in X^{p_1 ,..., p_N}$ is in such an irreducible representation if
%
%
\be
\sigma^{(ij)} A \; =\; 0
\label{eq:4.15}
\ee
for any $j \geq i$ and also satisfying $t^{(ij)} A =A$ if the $i$th and $j$th columns are of equal length, such that $p_i = p_j$. Such a representation is labelled $[ p_1 ,..., p_N ]$. Again, one can project onto this irreducible tensor subspace $X^{[p_1 ,...,p_N ]}$ from $X^{p_1 ,...,p_N}$ using the Young symmetriser ${\cal{Y}}_{[p_1 ,...,p_N ]}$.

The natural gauge transformation for this object is then given by
%
%
\begin{equation}
\delta A \;\; = \;\; {\cal{Y}}_{[ p_1 ,..., p_N ]} \circ \left( \, \sum^N_{i=1}\; d^{(i)} \alpha_{(i)}^{p_1 ,...,p_i -1 ,...,p_N } \, \right)
\label{eq:4.16}
\end{equation}
for any gauge parameters $\alpha_{(i)}^{p_1 ,..., p_i -1,..., p_N } \in X^{p_1 ,..., p_i -1,..., p_N }$. This just corresponds to the sum over $N$ tableaux of type $[ p_1 ,..., p_N ]$ with the $i$th term in the sum having a single partial derivative entered in the $i$th column.

The associated field strength $F$ is a type $[ p_1 +1 ,..., p_N +1]$ tensor defined by
%
%
\begin{equation}
F \;\; = \;\; {\cal{Y}}_{[ p_1 +1 ,..., p_N +1]} \circ \left( \, \prod_{i=1}^{N} \; d^{(i)} A \, \right) \;\; \equiv \;\; \prod_{i=1}^{N} \; d^{(i)} A
\label{eq:4.17}
\end{equation}
which is gauge invariant under ({\ref{eq:4.16}}). The identity in ({\ref{eq:4.17}}) follows from the observation that the composite exterior operator $\prod_{i=1}^{N} d^{(i)} : X^{[ p_1 ,..., p_N ]} \rightarrow X^{[ p_1 +1 ,..., p_N +1]}$ maps tableaux to tableaux even though individual exterior derivatives do not. The expression is unambiguous since all $d^{(i)}$ commute. The field strength ({\ref{eq:4.17}}) corresponds to a $[ p_1 +1 ,..., p_N +1]$ Young tableau with $N$ partial derivatives (one in each column). Gauge invariance then follows from $\delta F$ vanishing identically since at least one column must contain at least two partial derivatives. Again, rewriting the generalised Poincar\'{e} lemma in {\cite{key8}} allows any type $[ p_1 +1,..., p_N +1]$ tensor $F$ (satisfying $d^{(i)} F =0$ for all $i$) to be written as in ({\ref{eq:4.17}}) for some type $[ p_1 ,..., p_N ]$ potential $A$. The field strength $F$ also satisfies second Bianchi identities
%
%
\begin{equation}
d^{(i)} F\; =\; 0 
\label{eq:4.18}
\end{equation}
and the first Bianchi identities
%
%
\begin{equation}
\sigma^{(ij)} F\; =\; 0 
\label{eq:4.19}
\end{equation}
for any $j \geq i$. 

Considering the irreducible representations of $GL(D,{\mathbb{R}})$ above to be reducible representations of the $SO(D-1,1)$ Lorentz subgroup allows the construction of a gauge invariant action functional from which physical equations of motion can be obtained. This principle is perhaps best illustrated by the use of two simple, non-trivial examples.

Consider first the $N=1$ example of a one-form Maxwell gauge field $A_{\mu}$, viewed as a type $[1,0]$ tensor. The natural field equation for this model is the Maxwell equation $\partial^{\mu} \partial_{[ \mu} A_{\nu ]} =0$. This equation can be derived from a gauge invariant Lagrangian proportional to $A^{\mu} \partial^{\nu} \partial_{[ \mu} A_{\nu ]}$. The gauge invariant field equation factor $E_{\mu} = \partial^{\nu} \partial_{[ \mu} A_{\nu ]}$ corresponds to the trace of the type $[2,1]$ field strength tensor $F_{\mu\nu\rho} = 2\, \partial_{[ \mu} A_{\nu ],\rho}$. The derived field equations can then be written
%
%
\begin{equation}
\tau^{(12)} F\; =\; 0 
\label{eq:4.20}
\end{equation}

Now consider the $N=3$ example of a totally symmetric, third rank gauge field $\phi_{\mu\nu\rho}$, viewed as a type $[1,1,1,0]$ tensor. The type $[2,2,2,1]$ field strength $F = d^{(1)} d^{(2)} d^{(3)} d^{(4)} \phi$ is invariant under the most general gauge transformation $\delta \phi_{\mu\nu\rho} = 3\, \partial_{( \mu} \xi_{\nu\rho )}$ for second rank gauge parameter $\xi_{\mu\nu}$ which can be taken to be symmetric. A unique gauge invariant action is given by
%
%
\begin{equation}
{\cal{S}}^{[1,1,1,0]} \;\; =\;\; - \frac{1}{6} \int d^D x \; \phi^{\mu\nu\rho} E_{\mu\nu\rho}  
\label{eq:4.21}
\end{equation}
where
%
%
\begin{equation}
E_{\mu\nu\rho} \;\; =\;\; {\cal{Y}}_{[1,1,1,0]} \circ \left( F_{\alpha\mu\alpha\nu\beta\rho\beta} - \frac{1}{2} \eta_{\mu\nu} F_{\alpha\beta\alpha\beta\gamma\rho\gamma} \right) 
\label{eq:4.22}
\end{equation}
is the type $[1,1,1,0]$ field equation tensor, satisfying $\partial^{\mu} E_{\mu\nu\rho} \equiv 0$ identically. Consequently, ({\ref{eq:4.21}}) is invariant under the gauge transformation $\delta \phi_{\mu\nu\rho} = 3\, \partial_{( \mu} \xi_{\nu\rho )}$. The field equation derived from ({\ref{eq:4.21}}) is $E_{\mu\nu\rho} =0$ which implies ${\cal{Y}}_{[1,1,1,0]} \circ \left( F_{\alpha\mu\alpha\nu\beta\rho\beta} \right) =0$ in $D \not= 2$. In bi-form notation, this field equation is then written      
%
%
\begin{equation}
\sum \tau^{(ij)} \tau^{(k4)} F \;\; =\;\; 0 
\label{eq:4.23}
\end{equation}
where the sum is over three terms with $(ijk) \in \{ (123) , (231) , (312) \}$, corresponding to the $[1,1,1,0]$ Young symmetrisation. It should be noted that this field equation contains four derivatives of gauge field $\phi$ and consequently is somewhat unphysical in the sense that physical theories contain precisely two derivatives in their equations of motion. Examples of such higher derivative field equations occur in Weyl gravity and higher order gauge theories with Lagrangians quadratic in the curvature.  

For $N$ odd, the natural field equation for a general type $[ p_1 ,..., p_N ,0]$ gauge potential $A$, with type $[ p_1 +1 ,..., p_N +1,1 ]$ field strength $F$, is then given by
%
%
\begin{equation}
{\cal{Y}}_{[ p_1 ,..., p_N ,0]} \circ \left( \sum_{I \in S_N} \tau^{( i_1 i_2 )} ... \tau^{( i_N \, N+1)} F \right) \;\; =\;\; 0 
\label{eq:4.24}
\end{equation}
where the sum is on the labels $I = ( i_1 ... i_N )$ whose values vary over all permutations of the set $(1...N)$. The $(N+1)$th label is not included in the sum and the Young projection is onto an irreducible type $[ p_1 ,..., p_N ,0]$ tensor. For $N$ even, the field equation for a type $[ p_1 ,..., p_N ]$ gauge potential $A$, with type $[ p_1 +1 ,..., p_N +1 ]$ field strength $F$, is given by
%
%
\begin{equation}
{\cal{Y}}_{[ p_1 ,..., p_N ]} \circ \left( \sum_{I \in S_N} \tau^{( i_1 i_2 )} ... \tau^{( i_{N-1} i_N )} F \right) \;\; =\;\; 0 
\label{eq:4.25}
\end{equation}
where the sum here is on all the labels $I = ( i_1 ... i_N )$ whose values vary over all permutations of the set $(1...N)$. The  Young projection is then onto an irreducible type $[ p_1 ,..., p_N ]$ tensor.

These field equations can be derived from the gauge invariant action 
%
%
\begin{equation}
{\cal{S}}^{[ p_1 ,..., p_N ]} \;\; =\;\; - \left( \prod_{i=1}^{N} \frac{1}{p_i !} \right) \int d^D x \; A^{{\mu}^1_1 ...{\mu}^1_{p_1} ...{\mu}^i_1 ...{\mu}^i_{p_i} ...{\mu}^N_{p_N} } E_{{\mu}^1_1 ...{\mu}^1_{p_1} ...{\mu}^i_1 ...{\mu}^i_{p_i} ...{\mu}^N_{p_N}}  
\label{eq:4.26}
\end{equation}
in terms of the type $[ p_1 ,..., p_N ]$ gauge potential $A$ and some gauge invariant field equation tensor $E$ involving $N$ partial derivatives on $A$ for even $N$ (or $N+1$ derivatives for odd $N$). Gauge invariance of ({\ref{eq:4.26}}) necessarily implies that $E$ should satisfy the $N$ conservation conditions $\partial^{\mu^i_1} E_{{\mu}^1_1 ...{\mu}^1_{p_1} ...{\mu}^i_1 ...{\mu}^i_{p_i} ...{\mu}^N_{p_N}} \equiv 0$ identically for $i=1,...,N$. For $N$ even then the leading term in $E$ necessarily involves $N/2$ traces of the field strength $F$ of $A$ and is obtained by the $[ p_1 ,..., p_N ]$ Young symmetrisation of the sum over all permutations of $N$ labels of the term $F_{ {\mu}^1_1 ... {\mu}^1_{p_1 +1} ... {\mu}^N_{p_N +1} } \; \eta^{\mu^1_{1} \, \mu^2_{1}} ... \eta^{\mu^{N-1}_{1} \, \mu^N_{1}}$. The correction terms then consist of all further traces (appropriately symmetrised) with coefficients fixed by overall conservation of $E$. For $N$ odd one can consider the potential to be a type $[ p_1 ,..., p_N ,0]$ tensor of even order $N+1$ whose field strength $F$ is a type $[ p_1 +1,..., p_N +1 ,1]$ tensor. The construction of $E$ is then the same as for the even $N$ case. 

The notion of duality described in the previous sections could also be extended to give $2^N$ equivalent descriptions of a gauge theory with type $[ p_1 ,..., p_N ]$ gauge potential $A$. The construction is perhaps best illustrated by considering $A$ in physical gauge such that it transforms in the $SO(D-2)$-irreducible representation $\widehat{[ p_1 ,..., p_N ]}$. There are then $2^N$ dual descriptions in terms of each of the potentials $A$, $*^{(i)} A$, $*^{(i)} *^{(j)} A$,..., $*^{(1)} ... *^{(N)} A$. It must be understood however that not all of these dual descriptions are non-trivial. For example, in the type $[1,1]$ gravitational case then one has the set of four physical dual fields $( h, *h , {\tilde{*}}h , *{\tilde{*}}h )$ though only three of them $h$, $D = *h = t{\tilde{*}}h$ and $C=*{\tilde{*}}h$ are distinct due to the symmetry of $h$. In general the number of distinct dual descriptions would therefore depend on the particular representation $A$ is in.
                     

\vspace*{.6in}
{\bf{Note added:}}
Most of the new material here was presented in {\cite{key6}}. Subsequently, while this article was in preparation, the work {\cite{key14}} appeared which also develops a multi-form formulation of the results of {\cite{key5}} and so has considerable overlap with this paper.



\begin{thebibliography}{30}
\bibitem{keysezsun}E. Sezgin and P. Sundell, {\emph{JHEP}} {\textbf{0109}}, 025 (2001), {\texttt{hep-th/0107186}}; E. Sezgin and P. Sundell, {\emph{$7D$ bosonic higher spin gauge theory: symmetry algebra and linearised constraints}}, {\texttt{hep-th/0112100}}; E. Sezgin and P. Sundell, {\emph{Massless higher spins and holography}}, {\texttt{hep-th/0205131}}.
\bibitem{keylabmor}J.M.F. Labastida and T.R. Morris, {\emph{Phys. Lett.}} {\textbf{B180}}, 101 (1986).
\bibitem{keysieg}W. Siegel and B. Zwiebach, {\emph{Nucl. Phys.}} {\textbf{B282}}, 125 (1987); W. Siegel, {\emph{Nucl. Phys.}} {\textbf{B284}}, 632 (1987); W. Siegel, {\emph{Introduction to string field theory}}, chapter 4 (World Scientific, 1988), {\texttt{hep-th/0107094}}; N. Berkovits, {\emph{Phys. Lett.}} {\textbf{388B}}, 743 (1996), {\texttt{hep-th/9607070}}; W. Siegel, {\emph{Fields}}, chapter 12 (1999), {\texttt{hep-th/9912205}}.
\bibitem{keytsul}{\v{C}}. Burdik, {\emph{Mod. Phys. Lett.}} {\textbf{A16}}, 731 (2001), {\texttt{hep-th/0101201}}; {\v{C}}. Burdik, A. Pashnev and M. Tsulaia, {\emph{Nucl. Phys. Proc. Suppl.}} {\textbf{102}}, 285 (2001), {\texttt{hep-th/0103143}}; I.L. Buchbinder, A. Pashnev and M. Tsulaia, {\emph{Phys. Lett.}} {\textbf{B523}}, 338 (2001), {\texttt{hep-th/0109067}}; I.L. Buchbinder, A. Pashnev and M. Tsulaia, {\emph{Massless higher spin fields in the AdS background and BRST constructions for non-linear algebras}}, {\texttt{hep-th/0206026}}. 
\bibitem{key1}T.L. Curtright and P.G.O. Freund, {\emph{Nucl. Phys.}} {\textbf{B172}}, 413 (1980); T.L. Curtright, {\emph{Phys. Lett.}} {\textbf{B165}}, 304 (1985).
\bibitem{key2}C.M. Hull, {\emph{Nucl. Phys.}} {\textbf{B583}}, 237 (2000), {\texttt{hep-th/0004195}}.
\bibitem{key3}C.M. Hull, {\emph{Class. Quant. Grav.}} {\textbf{18}}, 3233 (2001), {\texttt{hep-th/0011171}}.
\bibitem{key4}C.M. Hull, {\emph{JHEP}} {\textbf{0012}}, 007 (2000), {\texttt{hep-th/0011215}}.
\bibitem{key5}C.M. Hull, {\emph{JHEP}} {\textbf{0109}}, 027 (2001), {\texttt{hep-th/0107149}}.
\bibitem{key6}P. de Medeiros, {\emph{Duality in non-perturbative quantum gravity}}, PhD thesis, Queen Mary, University of London {\texttt{[submitted 24/07/02]}}.
\bibitem{key7}M. Dubois-Violette and M. Henneaux, {\emph{Lett. Math. Phys.}} {\textbf{49}}, 245 (1999), {\texttt{math.QA/9907135}}.
\bibitem{key8}M. Dubois-Violette, {\emph{Lectures on differentials, generalized differentials and on some examples related to theoretical physics}}, {\texttt{math.QA/0005256}}.
\bibitem{key9}M. Dubois-Violette and M. Henneaux, {\emph{Tensor fields of mixed Young symmetry type and N-complexes}}, {\texttt{math.QA/0110088}}.
\bibitem{key10}M. Hamermesh, {\emph{Group theory and its application to physical problems}}, (Dover publications, 1989).
\bibitem{key11}J. Antonio Garcia and Bernard Knaepen, {\emph{Phys. Lett.}} {\textbf{B441}}, 198 (1998), {\texttt{hep-th/9807016}}.
\bibitem{key12}H. Casini, R. Montemayor and L.F. Urrutia, {\emph{Phys. Lett.}} {\textbf{B507}}, 336 (2001), {\texttt{hep-th/0102104}}.
\bibitem{key13}C.M. Hull, {\emph{Comm. Math. Phys.}} {\textbf{156}}, 245 (1993), {\texttt{hep-th/9211113}}.
\bibitem{key14}X. Bekaert and N. Boulanger, {\emph{Tensor gauge fields in arbitrary representations of $GL(D,{\mathbb{R}})$ : duality and Poincar\'{e} lemma}}, {\texttt{hep-th/0208058}}.
\end{thebibliography}
\end{document}